\documentclass[USenglish,twocolumn]{article}

\usepackage[utf8]{inputenc}%(only for the pdftex engine)
\usepackage[big]{dgruyter}

\newcommand{\eff}{_{\textrm{\tiny eff}}} % non-italic subscript "eff"
\newcommand{\subsun}{$_{\odot}$} % subscript solar mass symbol

\begin{document}

  \articletype{Research Article{\hfill}Open Access}

  \author*[1]{Ingrid Pelisoli}

\author[2]{S. O. Kepler}

\author[3]{Detlev Koester}

\affil[1]{Instituto de Física, Universidade Federal do Rio Grande do Sul, Porto Alegre, Rio Grande do Sul, 91501-900, Brazil, E-mail: ingrid.pelisoli@ufrgs.br}

\affil[2]{Instituto de Física, Universidade Federal do Rio Grande do Sul, Porto Alegre, Rio Grande do Sul, 91501-900, Brazil, E-mail: kepler@if.ufrgs.br}

\affil[3]{Institut für Theoretische Physik und Astrophysik, Kiel, Schleswig-Holstein, 24098, Germany, E-mail: koester@astrophysik.uni-kiel.de}

  \title{\huge Are sdAs helium core stars?}

  \runningtitle{Are sdAs He core stars?}

  %\subtitle{...}

  \begin{abstract}
{
Evolved stars with a helium core can be formed by non-conservative mass exchange interaction with a companion or by strong mass loss. Their masses are smaller than 0.5~M$_{\odot}$. In the database of the Sloan Digital Sky Survey (SDSS), there are several thousand stars which were classified by the pipeline as dwarf O, B and A stars. Considering the lifetimes of these classes on the main sequence, and their distance modulus at the SDSS bright saturation, if these were common main sequence stars, there would be a considerable population of young stars very far from the galactic disk. Their spectra are dominated by Balmer lines which suggest effective temperatures around 8\,000--10\,000~K. Several thousand have significant proper motions, indicative of distances smaller than 1~kpc. Many show surface gravity in intermediate values between main sequence and white dwarf, $4.75 < \log g < 6.5$, hence they have been called sdA stars. Their physical nature and evolutionary history remains a puzzle. We propose they are not H-core main sequence stars, but helium core stars and the outcomes of binary evolution. We report the discovery of two new extremely-low mass white dwarfs among the sdAs to support this statement.}
\end{abstract}
%  \classification[PACS]{}
 % \communicated{...}
 % \dedication{...}

  \journalname{Open Astronomy}
\DOI{DOI}
  \startpage{1}
  \received{..}
  \revised{..}
  \accepted{..}

  \journalyear{2014}
  \journalvolume{1}
%  \journalissue{1}

\maketitle
\section{Introduction}

The physical properties of main sequence stars can be reasonably inferred from their spectral type. The spectral classes from A to M show an increase in molecular bands, with hydrogen becoming less prominent, reflecting a decrease in effective temperature ($T_{\textrm{\tiny eff}}$). Similarly, the mass also decreases. As mass is the determinant factor on the lifetime of a star, hydrogen abundant main sequence stars (early-type) are short lived compared to cool, late-type stars. Dwarf A stars, in particular, have a main sequence lifetime shorter than 2~Gyr. Consequently, stars of type A and earlier should not be found in the Galactic halo, which is at least 10~Gyr old, unless they were accreted or recently formed.

Mining the Sloan Digital Sky Survey (SDSS), we were surprised to encounter thousands of objects classified by the pipeline as of type O, B and A. The SDSS bright saturation is about $g=14.5$, while the absolute magnitude of a dwarf A star is $M_g = 0$; thus, if indeed in the main sequence, these objects would mostly have to be in the halo, given their distance modulus $(g - M_g) > 14.5$ implying $d \gtrsim 8$~kpc and the fact that the SDSS operates mostly outside the disk ($b > 30^{\circ}$).

In Kepler et al. (2016), we fitted the spectra of these objects to spectral models derived from pure-hydrogen atmosphere models, and found thousands to show surface gravity with $\log g > 5.5$. Given the properties of a dwarf A star, its maximal $\log g$ is about 4.75 (see Romero et al. 2015 and references therein). White dwarfs resulting from single evolution, on the other hand, have a lower limit in $\log g$ of about 6.5--7.0 (e.g. Kilic et al. 2007). Objects with $4.5<\log g<6.5$ can result from binary evolution, as the hot subdwarf stars: binary interaction strips away the star's outer layers during core He burn, leaving a hot (T$_{\textrm{\tiny eff}}>$20\,000~K) lower mass (M$\sim$0.45~M$_{\odot}$) object. However, we found the objects to have $T\eff < 20\,000$~K, therefore they should not be core helium burning objects as the hot subdwarfs. We have dubbed this type of object subdwarf A stars (sdAs), referring to their sub-main sequence surface gravity and A-star-like spectra. This is nonetheless merely a spectroscopic classification: Their physical nature and evolutionary history remains an embarrassing puzzle.

A promising possibility was that these objects were new extremely-low mass white dwarfs (ELMs, M$\lesssim$0.3~M$_{\odot}$). For low-mass progenitors (M$\lesssim$2.0~M\subsun), the temperature for burning He is only reached after it has become degenerate. Therefore, if the outer layers of a low-mass star are stripped away before the He burning starts, a degenerate He core with a hydrogen atmosphere will be left: an ELM (see the ELM Survey: Brown et al. 2010, Kilic et al. 2011, Brown et al 2012, Kilic et al. 2012, Brown et al. 2013, Gianninas et al. 2015, Brown et al. 2016).
 
\begin{figure}
\includegraphics[width=0.5\textwidth]{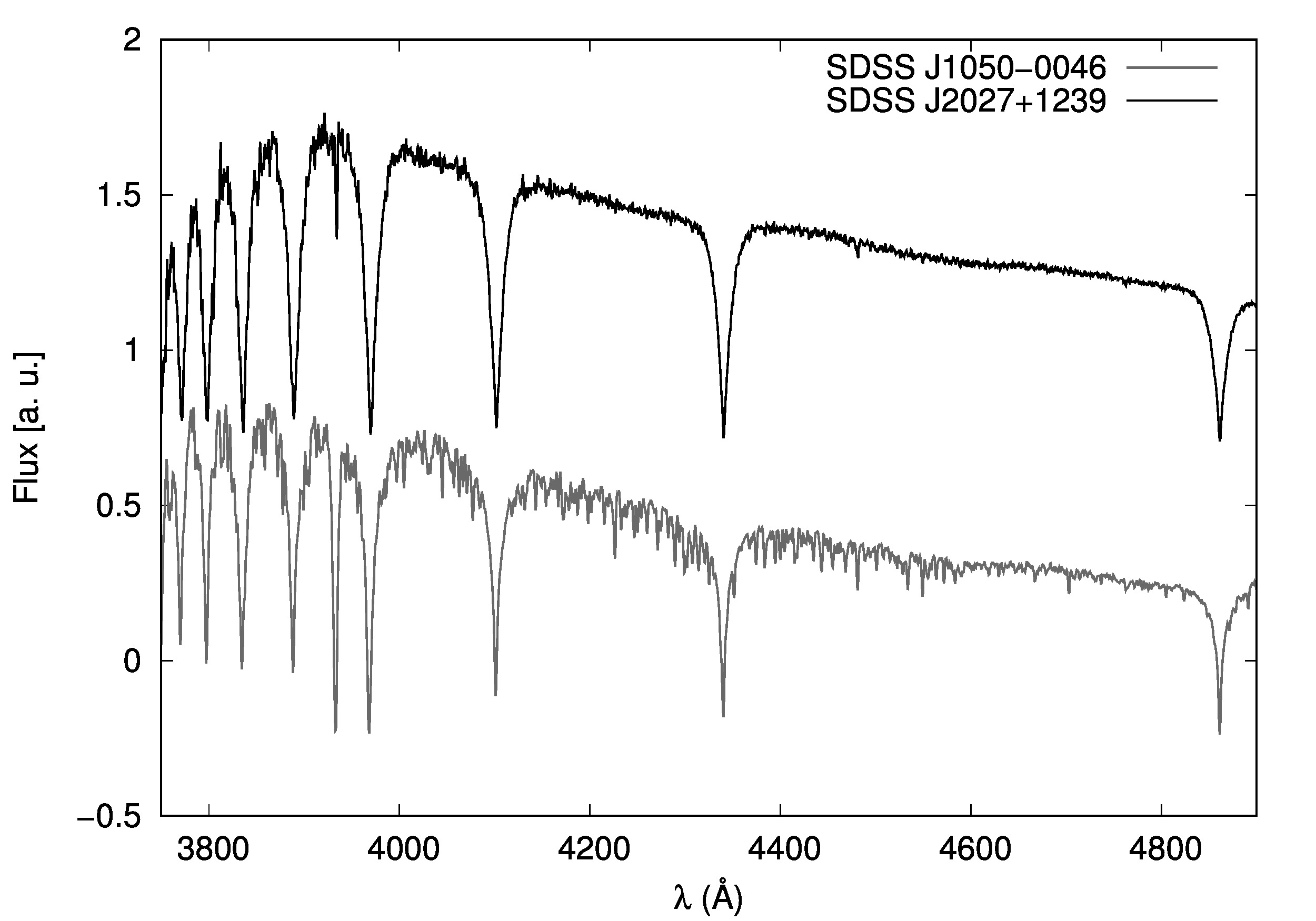}
\caption{Two sdA stars, SDSS~J105025.94-004655.5 (bottom) and SDSS~J202721.77+123942.7 (top). While SDSS~J1050-0046 shows lots of metallic lines, SDSS~J2027+1239 appears to have only a small amount of Ca and Mg.}
\label{fig2}
\end{figure}

Hermes et al. (2017) studied the sdAs that we published in Kepler et al. (2016), using radial 
velocity limits
obtained from SDSS subspectra, photometric colours, and reduced proper motions, and concluded that over 99 per cent of them are unlikely to be ELMs. Likewise, Brown et al. 2017 obtained follow-up time-resolved spectroscopy for five eclipsing systems and concluded they are not ELMs. 
They proposed these objects are metal-poor $M\sim1.2$~M\subsun{} main sequence stars with $M\sim0.8$~M\subsun{} companions, and suggested that the majority of sdAs are metal-poor A--F type stars. They argued that the $\log g$ of the sdAs was overestimated by $\sim 1$~dex on the surface gravities derived from pure hydrogen models, which is likely explained by metal line blanketing below 9000~K. As illustrated in Fig.~\ref{fig2}, some sdAs do show significant amount of metals in their spectra; however, the metals are almost undetectable in others. Brown et al. (2017) gives no explanation as to why or how these early-type stars are found in the halo.

\begin{figure}
\includegraphics[width=0.5\textwidth]{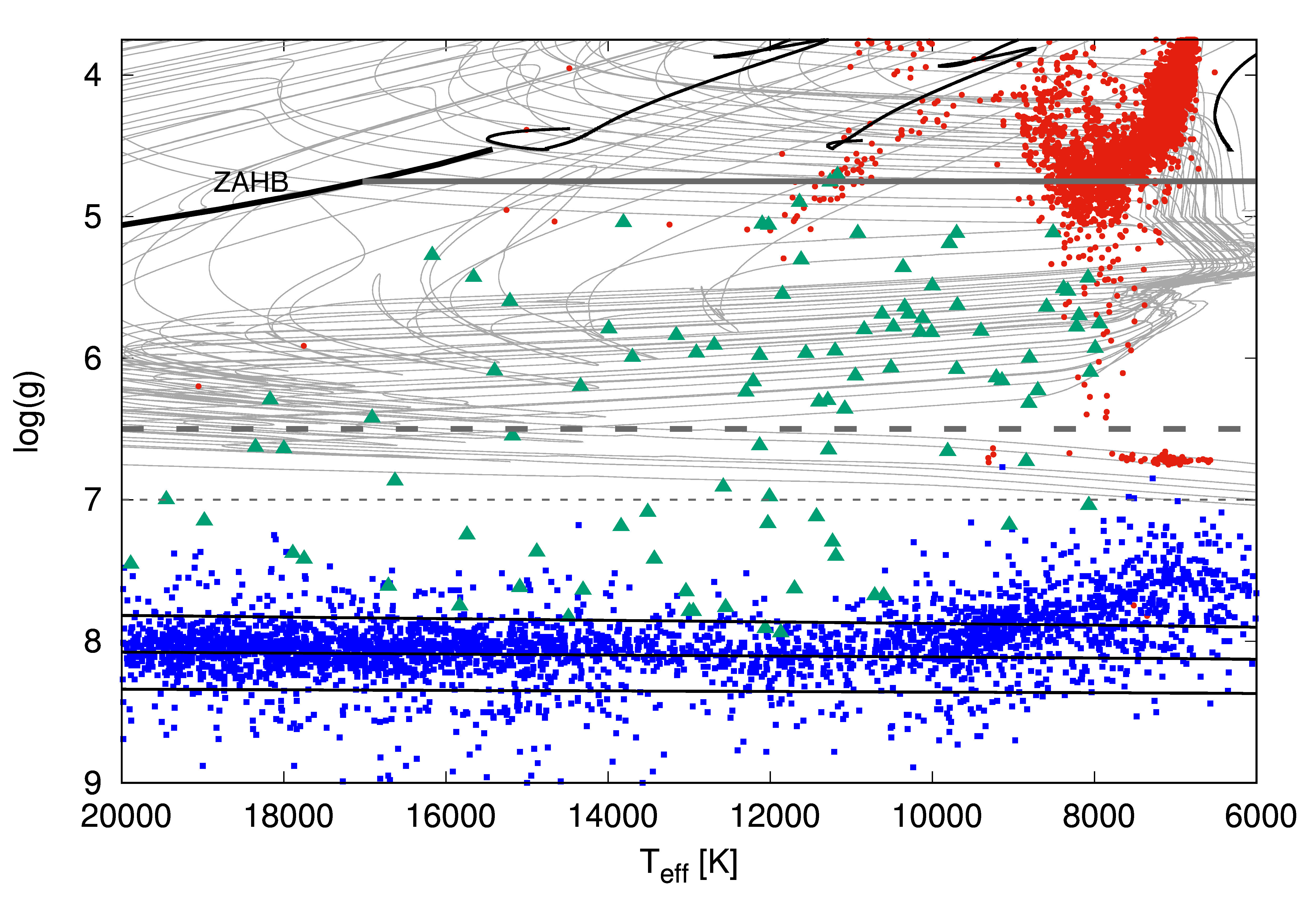}
\caption{Red dots show the fitted O, B, A type objects. The white dwarfs of Kepler et al. (2016) are shown as blue squares, and the known ELMs as green triangles for comparison. The zero-age horizontal branch (ZAHB), above which stars are burning He in the core, is indicated. The remaining black lines are single evolution models for 1.0, 2.0 and 3.0~M\subsun{} and Z=0.004 calculated with the {\sc LPCODE} (see Althaus et al. 2003 and references therein). The horizontal lines indicate the upper limit in $\log g$ for main sequence A stars (4.75) and the lower limit for white dwarf stars (6.5--7.0). The grey lines are the binary evolution models of Istrate et al. (2016), taking into account stellar rotation. Both the ELMs and the sdAs can be explained by these models.}
\label{fig1}
\end{figure}

An alternative that was overlooked by Brown et al. (2017) is that these objects are He-core stars and byproducts of binary interaction, including not only the ELMs, but the pre-ELMs, which have not reached the white dwarf cooling track yet, and blue straggler stars. Although stellar multiplicity is a function of mass, increasing from about 46 per cent for G-stars (Tokovinin 2014) to over 70 per cent for A stars (De Rosa et al. 2014), most stars with initial mass larger than 1.0~M\subsun{} are in multiple systems (Duchêne \& Kraus 2013), making this alternative very attractive. As shown in Fig. \ref{fig1}, the estimated $T\eff$ and $\log g$ of the sdAs are consistent with binary evolution models. Even though the time spent with $\log g = 5-6$ is ten times smaller than with $\log g = 6-7$ in the models of Istrate et al. (2016), the average luminosity is about a hundred times higher in the $\log = 5-6$ range, hence the objects are five magnitudes brighter. Assuming a spherical distribution, and limiting magnitudes of $g = 14.5$ (bright saturation in the SDSS) and $g = 20$ (faint limit detection), the detection volume for $\log = 5-6$ is a thousand times larger than the volume for  $\log = 6-7$. Combining these two factors, one should expect to find a hundred objects with $\log g = 5-6$ for each object with $\log g = 6-7$ in a magnitude-limited survey. Table 5 of Brown et al. (2016) lists 31 objects with $\log g = 6-7$, but only 44 with $\log g = 5-6$, about 85 per cent less than our estimate predicts, which is a consequence of their selection criteria.

Still, low ionisation potential metals can in fact contribute significantly to the electron pressure, so the issue raised by Brown et al. (2017) concerning the possible overestimate on the $\log g$ deserves attention. In Pelisoli et al. (2017), we have presented a brief analysis of the sdA population using a grid of solar metallicity models to account for the metal effect. In this work, we further analyse the sdA sample in the light of these new spectral models. Colours, proper motions, and galactic velocities are studied in order to access their possible nature. Analysing the SDSS subspectra, we find five new probable ELMs, two of which we confirm with our analysis of the SDSS radial velocities, and one also shows photometric variability in the Catalina Sky Survey (CSS) data. It seems that more than one evolution channel is needed to explain the sdA population. A definitive explanation of their nature and origin will help us to better understand not only stellar evolution, but also the formation of the halo.

\section{Methods}

The $55\,000+$ spectra of automatically classified O, B, A and white dwarf stars retrieved from the SDSS database were first fitted with  a grid of spectral models derived from pure hydrogen atmosphere models calculated using an updated version of the code described in Koester (2010). Objects with $\log g \geq 5.5$ were published in the SDSS DR12 white dwarf catalogue by Kepler et al. (2016) and were the first to be called sdAs. Both Hermes et al. (2017) and Brown et al. (2017) studied this DR12 sample reaching the conclusion that they are overwhelmingly not ELMs. The explanation of Brown et al. (2017) was an overestimate in $\log g$ resulting from the fact that pure hydrogen models ignore the effect of metal line blanketing. To account for that, we added metals, in solar abundances for simplicity, to our model atmosphere and synthetic spectra. Our grid covers $6\,000~\textrm{K} \leq T\eff \leq 40\,000$~K and $3.5 \leq \log g \leq 8.0$. The objects were fitted with this new grid first reported in Pelisoli et al. (2017).

While spectra are the most reliable way to estimate the physical properties of a star, the colours of an object alone can still be used as a complement and tell us something about its nature. The ultraviolet magnitudes, in particular, are very useful in identifying if the $T\eff$ of an object is high enough for it to be burning helium. We retrieved the far- and near-ultraviolet ($fuv$ and $nuv$) magnitudes from the Galaxy Evolution Explorer (GALEX) when available. Extinction correction was applied using the $E(B-V)$ value given on the GALEX catalogue, $R_{fuv} = 4.89$ and R$_{nuv}$ = 7.24 (Yuan et al. 2013).

Assuming the objects were main sequence stars, we estimated their distances $d$ by assuming a radius interpolated from solar-abundance values given the $T\eff$ of the object. The distance was calculated from the solid angle, which is estimated in a photometric fit to the SDSS $ugriz$ magnitudes and the GALEX $fuv$ and $nuv$ magnitudes. Given the galactic latitude $b$, we estimated the distance from the disk $Z$ as $d \sin(b)$.

We studied the proper motion of the O, B, A type objects using a reduced proper motion diagram (e.g. Gentile-Fusillo et al. 2015), where the reduced proper motion is given by:
\begin{eqnarray}
H_g &=& g_0 + 5\log(\mu[''/yr]) + 5.
\end{eqnarray}
It can be interpreted as a proxy for the absolute magnitude: the higher the reduced proper motion, the fainter the object. We used the proper motions of Munn et al. (2004) and Munn et al. (2014), given in the SDSS tables. They were obtained combining the data from the U.S. Naval Observatory (USNO) and the SDSS. We only show in the plot objects with reliable proper motion, namely with the following characteristics:
\begin{itemize}
\item proper motion > $3 \sigma_{\textrm{ppm}}$;
\item distance to nearest neighbour with $g>22.0$ larger than 5'';
\item only one matched object in the USNO catalogue;
\item at least four detections in the USNO catalogue plates;
\item RMS residual for the proper motion fit in right ascension smaller than 500.0;
\item RMS residual for the proper motion fit in declination smaller than 500.0.
\end{itemize}
Typical errors for the whole sample are 2--4~mas/yr; for the reliable proper motion sample this goes down to 0.5~mas/yr. For objects with a good proper motion, we have also evaluated the galactic velocities $U$, $V$, and $W$ following Johnson \& Soderblom (1987), 
with the radial velocities we derived from the spectra, assuming both a main sequence and an ELM radius. %%not accepted here

To search for binaries in the sample, we have used the SDSS subspectra. Each final SDSS spectrum is composed by multiple spectra, usually three, with $\sim 15$~min exposure time. The signal-to-noise ration (S/R) of the subspectra is almost always below ten, so while conclusions can hardly be made based solely on the SDSS subspectra, they can be used to probe for possible variations suggesting the need for a follow-up. Our approach is similar to that of Badenes \& Maoz (2012) and Hermes et al. (2017). We normalise each subspectrum by the continuum, which is estimated by fitting a linear function between each of the Balmer lines, and then fit each of the lines (up to H8) to a Gaussian profile. The obtained redshift to the line centre is used to estimate a radial velocity for each line. The final radial velocity for the given subspectrum is assumed to be the average velocity, with the error estimated by the standard deviation.

We were able to obtain a fit to 80 per cent of the spectra in the O, B, A sample. We then evaluated the $\Delta V$ between the maximal and the minimal estimated radial velocities, considering only estimates with an error smaller than 100~km/s. Badenes \& Maoz (2012) suggest that follow-up is needed to reach conclusions on objects that show $\Delta V < 200$~km/s, so we restrict further analysis to 14 objects showing $\Delta V > 200$~km/s. We used the Period04 software (Lenz \& Breger 2005) to estimate the orbital period by doing a Fourier transform and finding the orbital solution with the smallest residuals.

\section{Results}

\subsection{Spectral fits}

\begin{figure}
\includegraphics[angle=-90,width=0.5\textwidth]{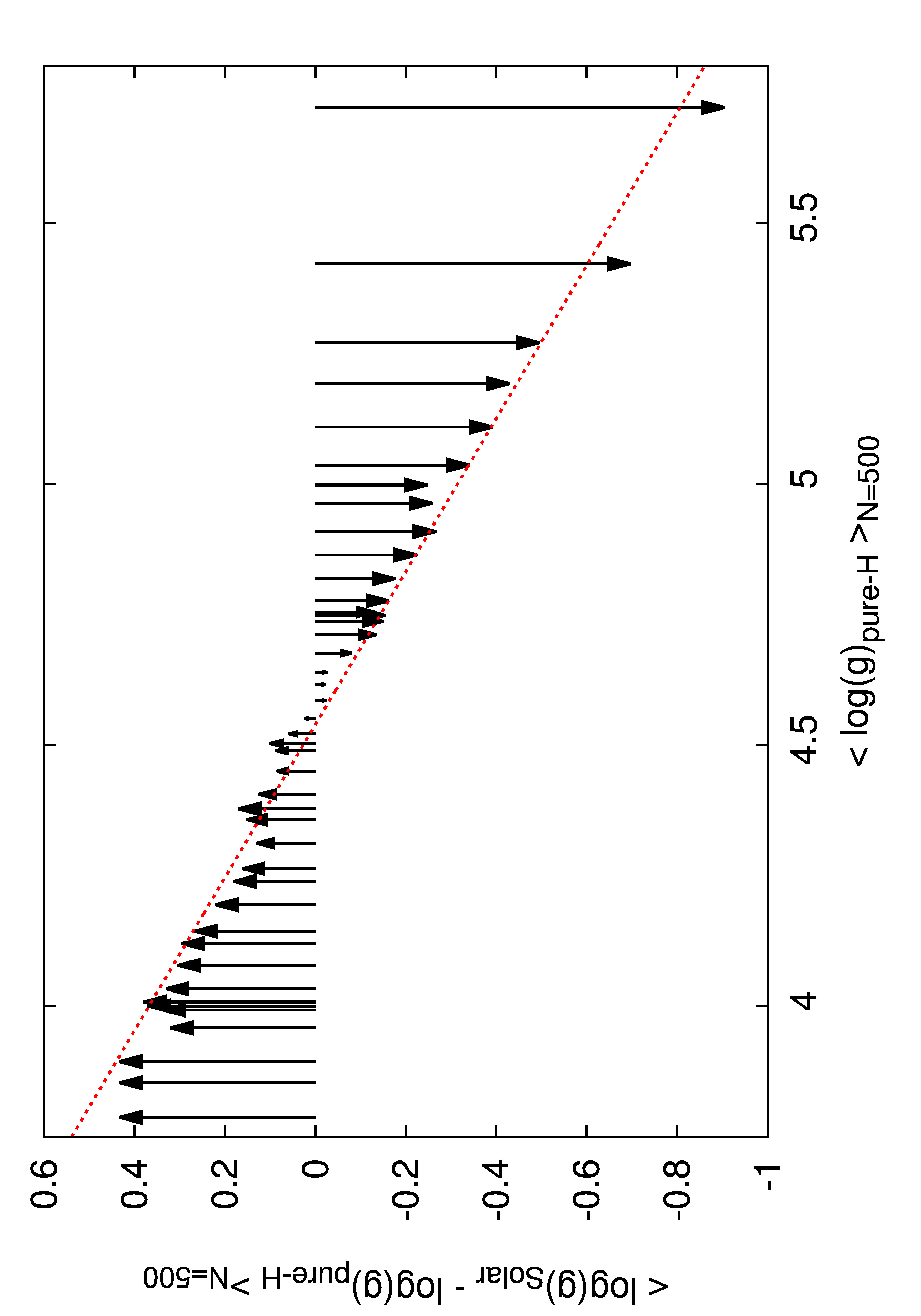}
\caption{Shift in $\log~g$ with the addition of metals in solar abundances as a function of the $\log~g$ given by the pure-H models. Values were averaged over 500 objects sorted by $\log~g$. The shifts are well described by a linear fit $\Delta \log~g = -0.68(0.01)\,\log~g_{\textrm{pure-H}} + 3.10(0.06)$, shown as a red dashed line. The pure-H values are almost 1.0~dex higher than the solar abundance values above $\log~g = 5.5$. This is a similar result to the obtained by Brown et al. (2017) when when fitting pure hydrogen model to synthetic main-sequence spectra.}
\label{logg_s}
\end{figure}

\begin{figure}
\includegraphics[width=0.5\textwidth]{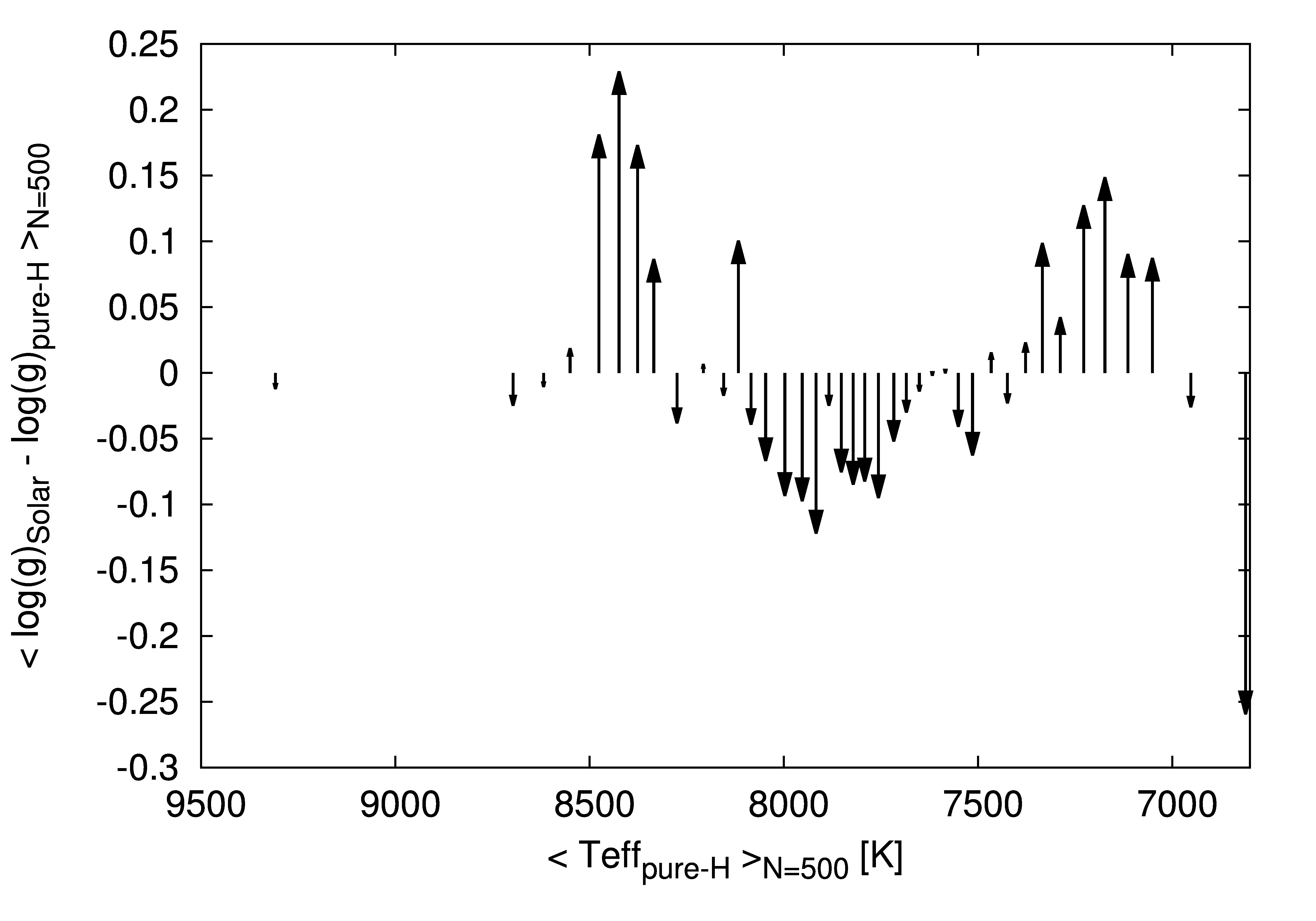}
\caption{Change in log g when metals were added to the models as a function of the effective temperature of the pure-H models. The $T\eff$ and the change in $\log g$ were averaged over 500 objects, sorted by $T\eff$. The apparent puzzling behaviour is a consequence of the systematic effect found for as a function of $\log~g_{\textrm{pure-H}}$, which implies a correlation also in $T\eff$, depending on how each range of $\log g$ is sampled in each bin of $T\eff$, as discussed in the text.}
\label{teff_s}
\end{figure}

The shifts in $\log g$ and in $T\eff$ when going from a pure-hydrogen model to a solar abundance model are shown in Figs. \ref{logg_s} and \ref{teff_s}. They were averaged between 500 objects, with the sample sorted according to $\log g$ or $T\eff$, respectivelly. Only objects with $T\eff$ differing by less than 500~K were taken into account, to avoid contamination by objects suffering from hot-cool solution degeneracy. We find that the addition of metals does not cause a constant shift in $\log g$ as suggested by Brown et al. (2017). The shift behaves linearly, with $\log g < 4.5$ objects showing an upward correction and $\log g > 4.5$ showing a downward correction. Above $\log g = 5.5$, where the sdAs of Kepler et al. (2016) are, about -1.0~dex is indeed the shift, as found by Brown et al. (2017). However, as the shift can go either way, even though the addition of metals solves the $\log g$ discrepancy for a few objects, others are raised above the $\log = 5.0$ limit, and still can not be explained by single evolution, even when metals are taken into account. %different here

This systematic trend also reflects on the dependence of the $\log g$ change with $T\eff$, shown on Fig. \ref{teff_s}. At $T\eff \sim 8\,500$~K, there are objects spanning all the $\log g$ range (see Fig. \ref{fig1}), but a prevalence of objects with lower $\log g$, which have an upward correction. Hence the same upward correction is seen in this $T\eff$ range. Between $7\,500 - 8\,000$~K, a gap in the lower $\log g$ objects can be seen on Fig. \ref{fig1}, which moves the correction downwards. Finally, below $T\eff \sim 7\,500$~K, most objects show $\log~g \leq 4.5$, so the correction moves upwards again. Close to the cool border of $T\eff$, most objects are also close to the lower border in $\log~g$, which is 3.75 for the pure-hydrogen models and 3.5 for the solar abundance models, implying on an average difference of 0.25. There can of course be differences in metallicity and errors in the determination, so individual objects can somewhat obscure these trends. 

The solar abundance solutions put most of the 2\,443 sdAs published in by Kepler et al. (2016) in the main sequence range, with the exception of 39 objects with still show $\log~g \geq 5.0$. Only seven out of those maintain $\log~g \geq 5.5$ in the solar abundance models, two of them  were published on the ELM Survey, (SDSS~J074615.83+392203.1 in Brown et al. 2012, and SDSS~J091709.55+463821.7 in Gianninas et al. 2015). However, given that the change in $\log g$ can also be upward, other objects are raised above the main sequence $\log g$ limit. We find 1\,952 objects to show $5.0 \leq \log(g) < 7.0$ and $T\eff < 20\,000$~K; out of those, 492 show $\log g > 5.5$. %%rejected here

\subsection{Distance and Velocities}

\begin{figure}
\includegraphics[width=0.5\textwidth]{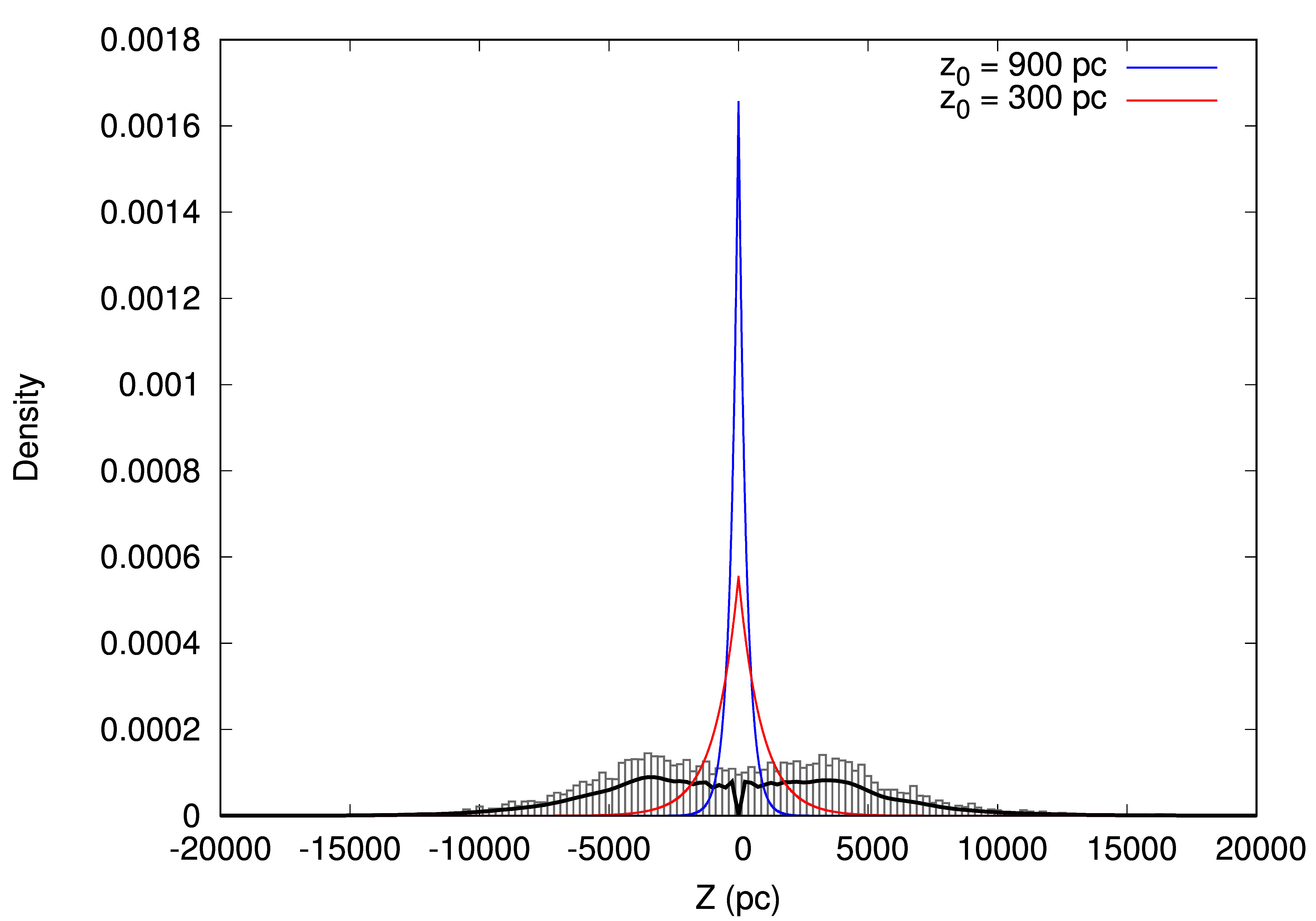}
\caption{The distance to the disk of the stars classified as O, B and A, assuming a main sequence radius. The histogram is given as $N/N_{\textrm{total}}$; the solid black line is calculated assuming each point as a Gaussian with standard deviation of 0.1~Z. The red line is an exponential thin disk model assuming $Z_0 = 300$~pc, while the blue line is a thick disk model with $Z_0 = 900$~pc. All functions are normalised. It is clear that, if indeed main sequence objects, these stars are not consistent with a disk distribution, but would rather have to be in the halo.}
\label{dist}
\end{figure}

Fig. \ref{dist} shows a histogram of the density $N/N_{\textrm{total}}$ given the estimated distances from the disk for the sample of O, B and A stars assuming they have main sequence radii. Exponential functions describing a thin and thick disk with the scaleheights given by Bland-Hawthorn \& Gerhard (2017) are shown as a comparison. It is clear that, when a main sequence radius is assumed, the sdA distribution extends much further than the disk, to distances up to 10~kpc.

\begin{figure}
\includegraphics[width=0.5\textwidth]{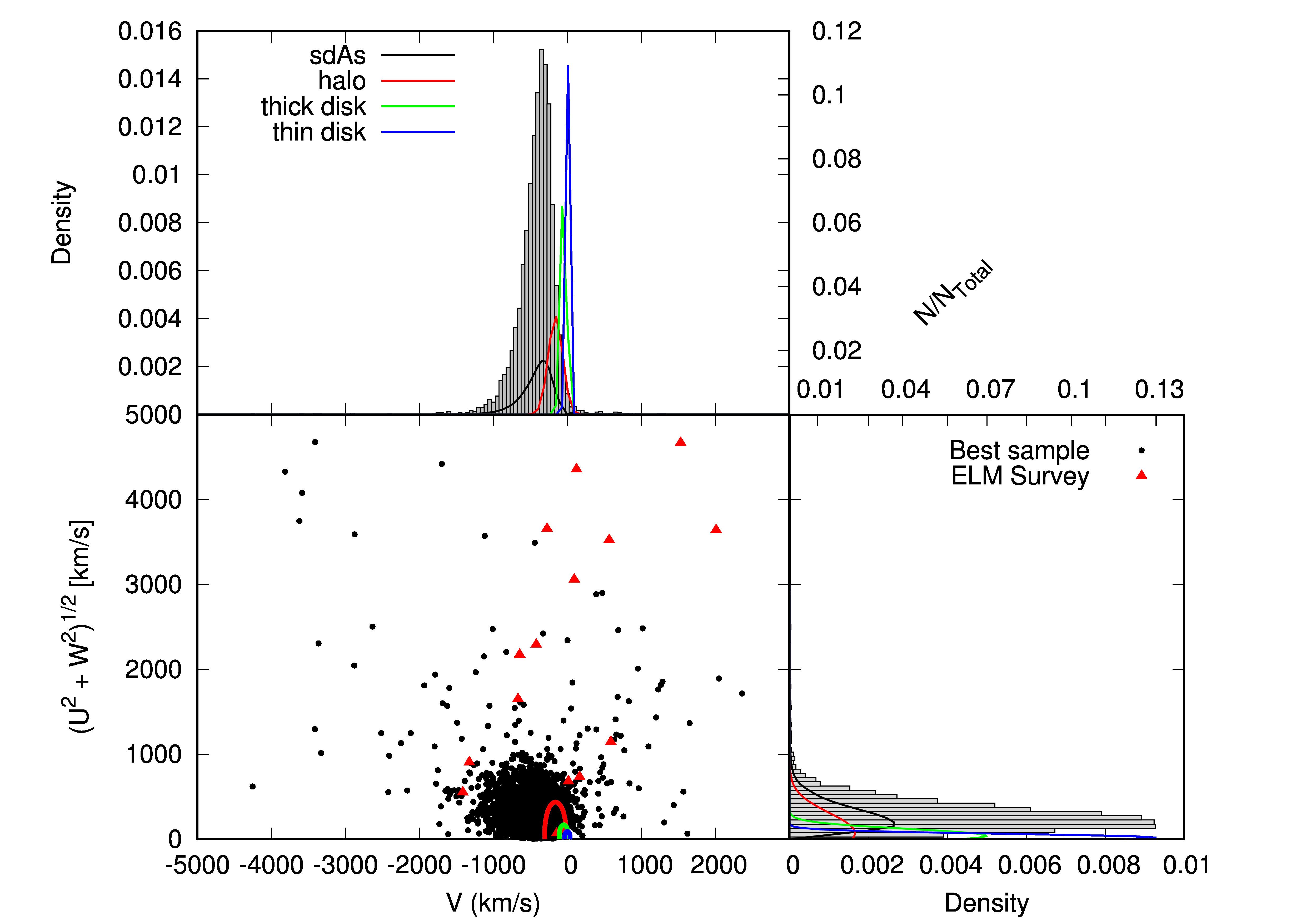}
\caption{Toomre diagram of the objects in our sample, assuming a main sequence radius. The velocities the objects in the ELM survey would have if main sequence radii were assumed are shown for comparison. Density plots are shown to left and on top. The ellipses indicate the 3-$\sigma$ values for halo (red), thick disk (green) and thin disk (blue) according to Kordopatis et al. (2011).}
\label{uvw}
\end{figure}

A similar result occurs when the Galactic velocities $U, V, W$ are estimated. Fig. \ref{uvw} shows the velocities estimated assuming the main sequence radius. Ellipses with the 3-$\sigma$ value for the thin disk, thick disk and halo, according to Kordopatis et al. (2011), are shown as a comparison. Again, the objects seem to reach velocities much higher than the disk distribution, and even than the halo distribution. In fact, over 30 per cent of the stars have velocities more than 4-$\sigma$ above the halo mean velocity dispersion when a main sequence radius is assumed. Even if we assume the distance is systematically 10~per cent smaller than our estimate, over 20 per cent of the objects show velocities above 4-$\sigma$. The statistical uncertainty is however set to zero when calculating the tangential velocities, so the identification of individual significant outliers requires caution. Considering the sample as whole though, it follows that metal-poor A--F main sequence is probably a too simplistic explanation for these objects. At the very least, they must be in a binary to account for the high velocities, which could be due to orbital motion rather than Galactic motion.

\subsection{Reduced proper motion}

\begin{figure}
\includegraphics[angle=-90,width=0.5\textwidth]{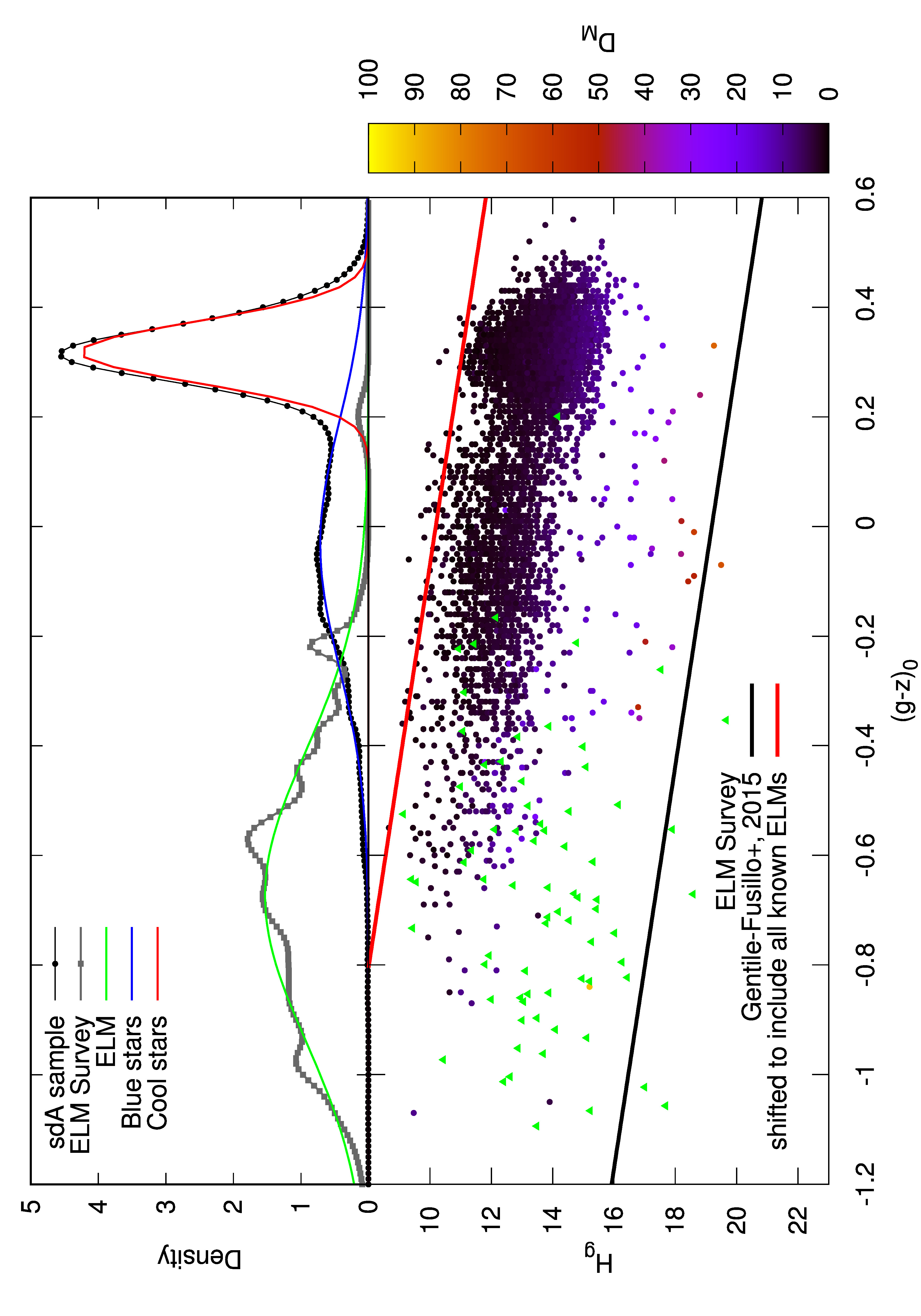}
\caption{$H_g \times (g-z)_0$ diagram (see e.g. Gentile-Fusillo et al. 2015), with the objects in sample B colour coded according to their Mahalonobis distance to the halo given a main sequence radius. Known ELMs are shown as green triangles for comparison. The top plot shows the densities assuming each object as a Gaussian to account for the uncertainty; it becomes clear that there are two populations of objects within the sdA sample. The suggested limit for white dwarf detection with probability equal to 1.0 given by Gentile-Fusillo et al. (2015) is indicated as a black solid line. Most known ELMs, due to their larger radius implying a smaller reduced proper motion, since they can be detected at larger distances, are not below the white dwarf limit. A reference line, dislocating the white dwarf limit to include all known ELMs is shown as a red dashed line. Most O, B and A stars are also below such line.}
\label{ppm}
\end{figure}

The reduced proper motion for the O, B, and A stars is shown in Fig. \ref{ppm}. It suggests that most of these objects have, in average, $H_g$ lower than the estimated for known ELMs. However, their reduced proper motion is mostly consistent with a tentative limit based on Gentile-Fusillo et al. (2015), but dislocated to 
include
all ELMs, suggesting the objects might have similar absolute magnitude, and thus similar radii, to the known ELMs. This limit is given by
\begin{eqnarray}
H_g= 2.72(g-z)_0 + 16.09.
\end{eqnarray}
The objects are colour coded by their Mahalonobis distance $D_M$ to the halo when a main sequence radius is assumed. The Mahalonobis distance is given by
\begin{eqnarray}
D_M = \sqrt{ \frac{\left( U - \langle U \rangle \right)^2}{\sigma^2_U} + \frac{\left( V - \langle V \rangle \right)^2}{\sigma^2_V} + \frac{\left( W - \langle W \rangle \right)^2}{\sigma^2_W}},
\label{mahalonobis}
\end{eqnarray}
where we have assumed the values of Kordopatis et al. (2011) for the halo mean velocities and dispersions.

This diagram is very enlightening when we look at the $(g-z)_0$ colour in terms of density. It is evident that there are two different populations within the sample: one to the red limit of the diagram and another in an intermediary region. While the distribution of the red population has no intersection with the known ELMs, the distribution resulting from the blue population shares colour properties with the known ELMs. This is a clear indication that more than one evolutionary channel is needed to explain the nature of these objects. The red distribution contains about 60 per cent of the sample. Most of these objects ($\sim 97$~per cent) are cooler than 8\,000~K and show $\log g < 4.75$, implying they may be low metallicity F stars or other late-type objects, which can be found in the halo. The blue population, on the other hand, contains about 40 per cent of the sample and most of the objects hotter than 8\,000~K (A-type and earlier) and with higher $\log g$. These early-type stars can not easily be explained as halo objects, since their life time in the main sequence is much smaller than the age of the halo. This population  probably consists of binaries, such as blue stragglers, and He-core objects, such as blue horizontal-branch stars (BHBs), as previous studies in the literature have found (e. g. Preston, Beers \& Shectman 1994, Clewley et al. 2004, Brown et al. 2008, Xue et al. 2008), or pre-ELMs, and ELMs. However, there can also be a contribution from extragalactic stars accreted onto our Galaxy, as previously suggested by e.g. Rodgers et al.(1981), Lance (1988), and  Preston, Beers \& Shectman (1994).

\subsection{UV colours}

\begin{figure}
\includegraphics[width=0.5\textwidth]{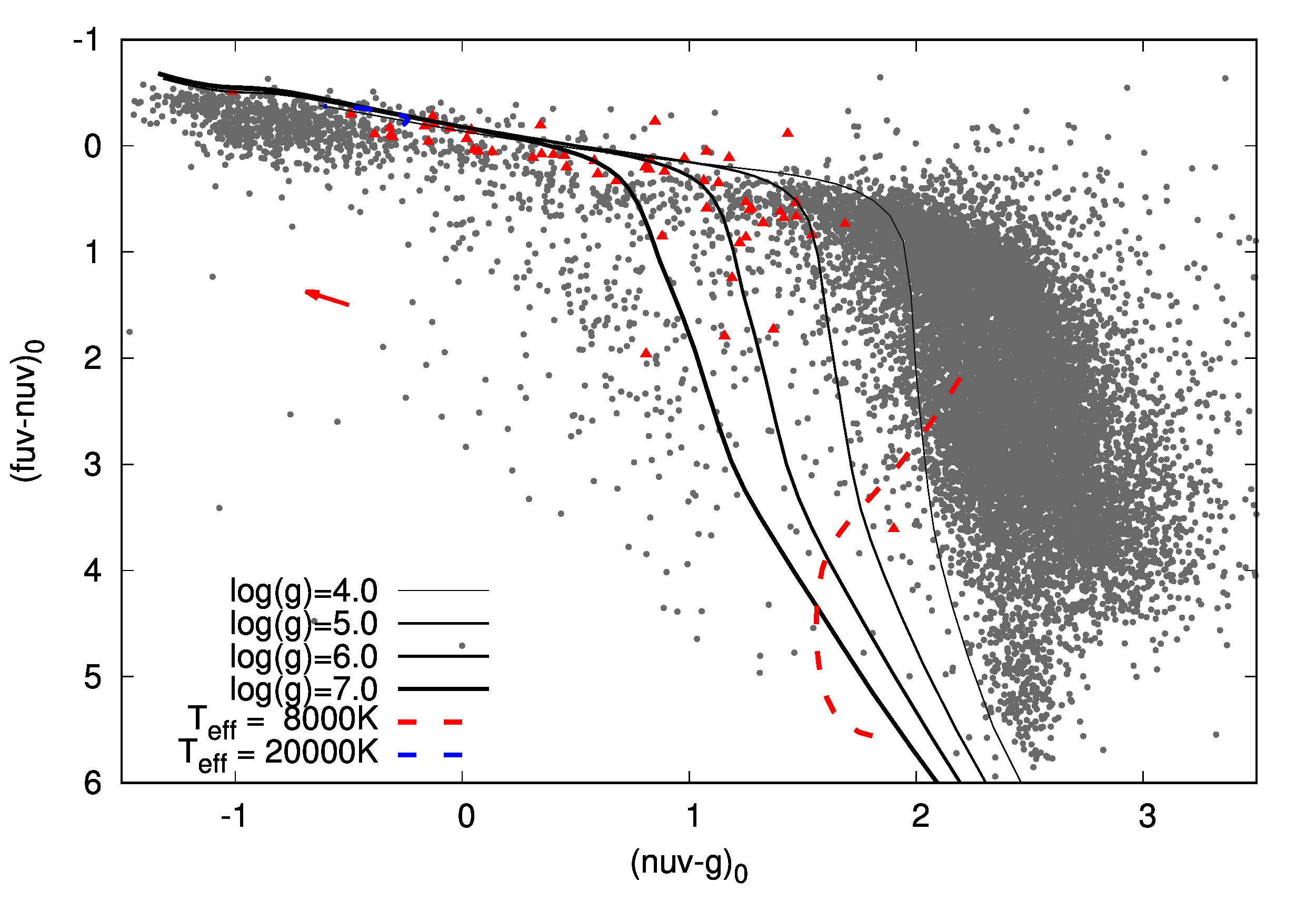}
\caption{Diagram showing the $(fuv-nuv)_0$ and $(nuv-g)_0$ colours. Grey dots are the O, B, and A objects, and red triangles are the known ELMs shown for comparison. The red arrow indicates the average reddening correction vector. The indicated models were obtained from our pure-hydrogen spectral models.}
\label{fnuvg}
\end{figure}

Fig. \ref{fnuvg} shows a $(fuv-nuv)_0 \times (nuv-g)_0$	for the O, B, and A stars and known ELMs for comparison. This diagram is especially useful in identifying if the objects can be hot subdwarfs in binaries. Hot subdwarf stars have similar flux in the optical region to main sequence stars of type F, G, and K, so that if they are in a binary with one of these types of stars, the combined spectrum will appear to have an intermediary $\log g$, but a lower temperature, similar to what is found for the sdAs. We find that almost all the objects, with a 0.5 per cent exception, do not have significant flux in the UV, showing $(nuv-g)_0 < -0.4$, which rules out that these objects can be explained as sdOB + FGK binaries.

\subsection{SDSS Radial Velocities}

\begin{figure}
\includegraphics[width=0.5\textwidth]{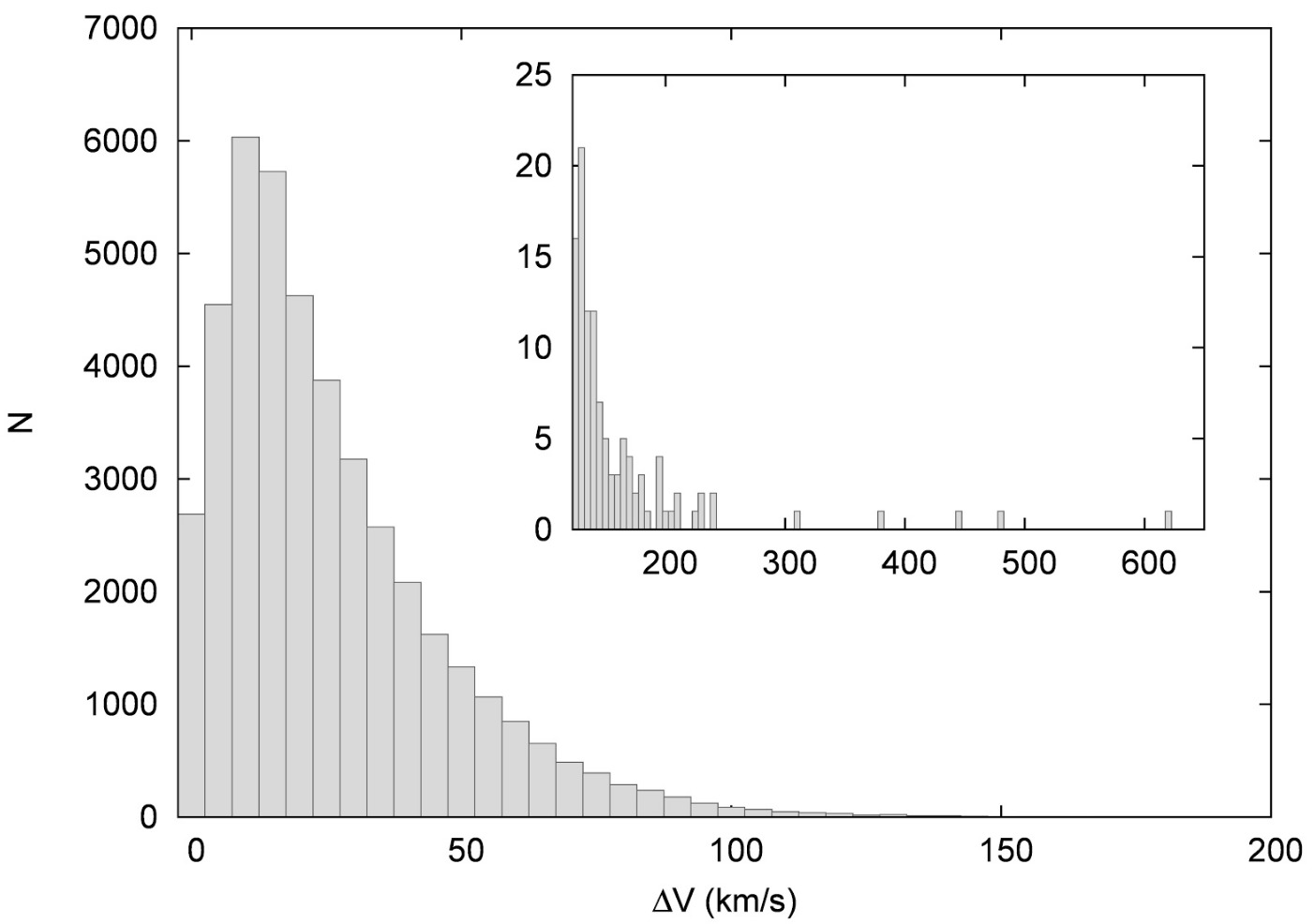}
\caption{Histogram showing the obtained amplitude for all analysed SDSS spectra. Most show no significant amplitude, but over 300 indicate an amplitude between subspectra larger than 100~km/s, 14 larger than 200~km/s.}
\label{histRV}
\end{figure}

\begin{figure}
\includegraphics[angle=-90,width=0.5\textwidth]{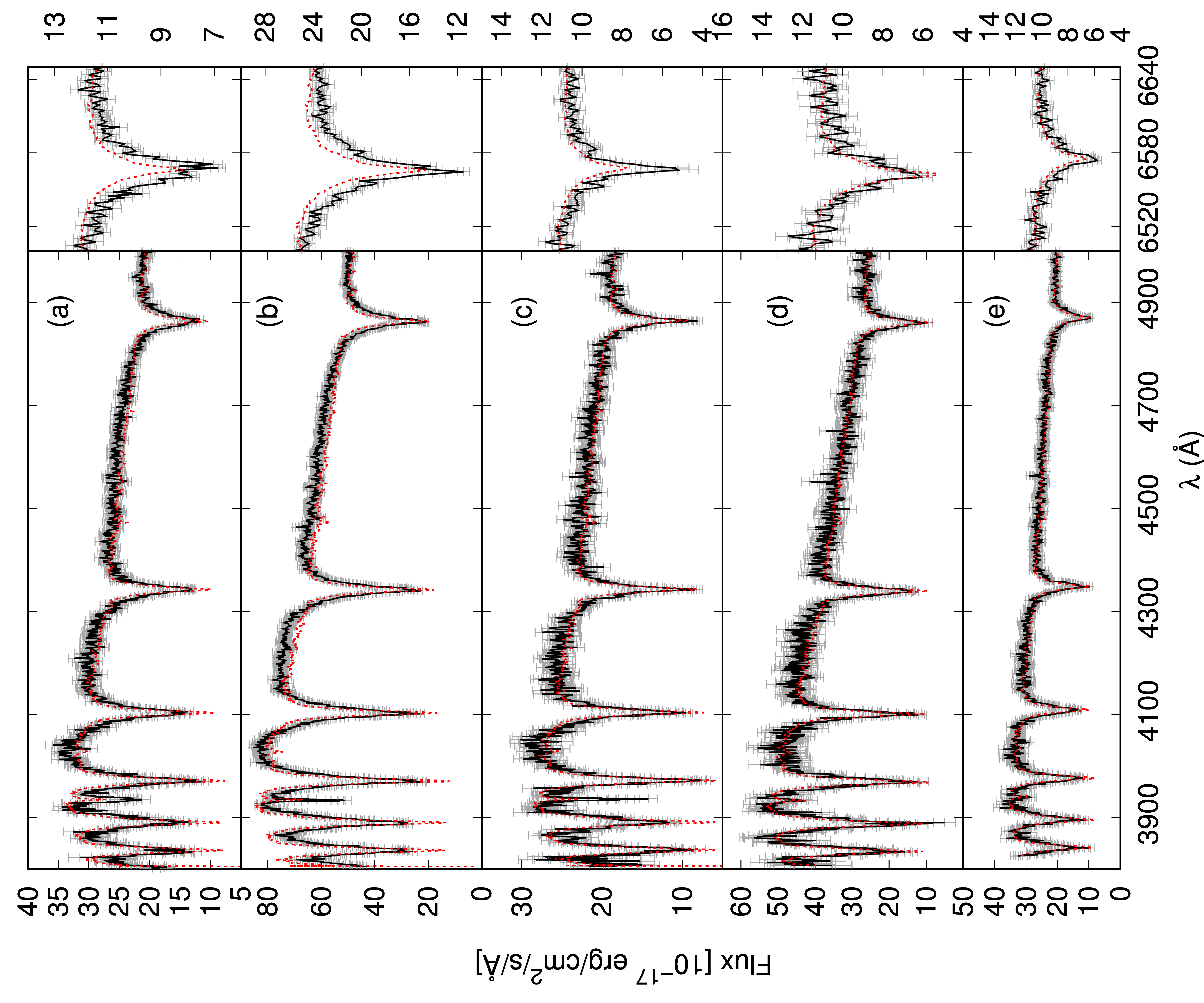}
\caption{Spectra for the five objects showing $\Delta V > 200$~km/s (solid black line). The SDSS template is shown as a dashed red line for comparison. For the object with two spectra, the highest $S/R$ spectrum is shown.}
\label{SDSS_ELMs_specs}
\end{figure}

Fig. \ref{histRV} shows a histogram of the estimated radial velocity amplitude $\Delta V$ from the SDSS subspectra. Most spectra show $\Delta V < 100$~km/s, with 334 having $\Delta V > 100$~km/s. Out of those, 14 show $\Delta V > 200$~km/s. Two of these objects were previously published in the ELM Survey, namely SDSS~J123800.09+194631.4 (Brown et al. 2013) and SDSS~J082511.90+115236.4 (Kilic et al. 2012). Three are hot subdwarf stars showing $T\eff > 20\,000$~K, which are also commonly found in binaries (SDSS~J141558.19-022714.3, SDSS~J163205.75+172241.3, and SDSS~J211651.95-003328.5). Two show $\log g > 7.0$ and are probably double degenerate systems (SDSS~J095157.78+290341.5 and SDSS~J132232.12+641545.8). One is a known CV (SDSS~J152020.40-000948.3) identified by its colours by Gentile-Fusillo et al. (2015). The remaining six spectra belong to five objects. The spectra are shown in Fig. \ref{SDSS_ELMs_specs}. Their atmospheric parameters are shown in Table \ref{tab_SDSS_ELMs}, for solar abundance models, and in Table \ref{tab_SDSS_ELMs2}, for pure-hydrogen atmosphere.

\begin{table*}
	\centering
	\caption{Atmospheric and orbital parameters obtained for the objects shown in Figs. \ref{SDSS_ELMs_specs} and \ref{SDSS_ELMs_orb}, assuming the solar abundance models. Quoted uncertainties in our values of $T\eff$ and $\log g$ are formal fit errors. The external uncertainties in the models are much larger, of about 5–10 per cent in $T\eff$ and 0.25 dex in $\log g$. The orbital parameters are for the best solution, but some objects might need follow-up (see text for discussion). The secondary mass $M_2$ is the minimal mass assuming an edge-on orbit.}
	\label{tab_SDSS_ELMs}
	\begin{tabular}{ccccccccccc}
		\hline
		 & SDSS~J & $g$ & $T\eff$ & $\log(g)$ & $P$ (h) & $K$ (km/s) & $R^2$ & $M_2$ ($M$\subsun) & $\mathcal{T}_{\textrm{merge}}$ (Gyr)\\
		\hline
		(a) & 104826.86-000056.7 & 18.39 &  8\,508(17) & 5.861(0.068) & 2.9 & 246 & 0.88 & 0.32 & 2.7 \\
		(b) & 120616.93+115936.2 & 17.37 &  8\,869(12) & 5.092(0.050) & 6.4 & 220 & 1.00 & 0.50 & 16 \\
                (c) & 045947.40-040035.2 & 19.62 &  8\,182(21) & 4.804(0.113) &  61 &  53 & 0.82 & 0.18 & 11280\\
		(d) & 171906.23+254142.3 & 19.13 &  8\,566(41) & 4.126(0.128) &  13 & 197 & 1.00 & 0.75 & 69\\
		(e) & 122911.49-003814.4 & 18.27 &  8\,020(22) & 4.657(0.128) & - & - & - & - & - \\
		\hline
	\end{tabular}
\end{table*}

\begin{table*}
	\centering
	\caption{Pure-hydrogen atmosphere spectral parameters for the objects shown in Table \ref{tab_SDSS_ELMs}. As before, the uncertainties are of about 5–10 per
cent in $T\eff$ and 0.25 dex in $\log g$.}
	\label{tab_SDSS_ELMs2}
	\begin{tabular}{cccc}
		\hline
		 & SDSS~J & $T\eff$ & $\log(g)$ \\
		\hline
		(a) & 104826.86-000056.7 & 8571 & 6.269\\
		(b) & 120616.93+115936.2 & 8861 & 5.308\\
                (c) & 045947.40-040035.2 & 8153 & 4.815\\
		(d) & 171906.23+254142.3 & 11288 & 4.500\\
		(e) & 122911.49-003814.4 & 8083 & 5.339\\
		\hline
	\end{tabular}
\end{table*}

\begin{figure}
\includegraphics[angle=-90,width=0.5\textwidth]{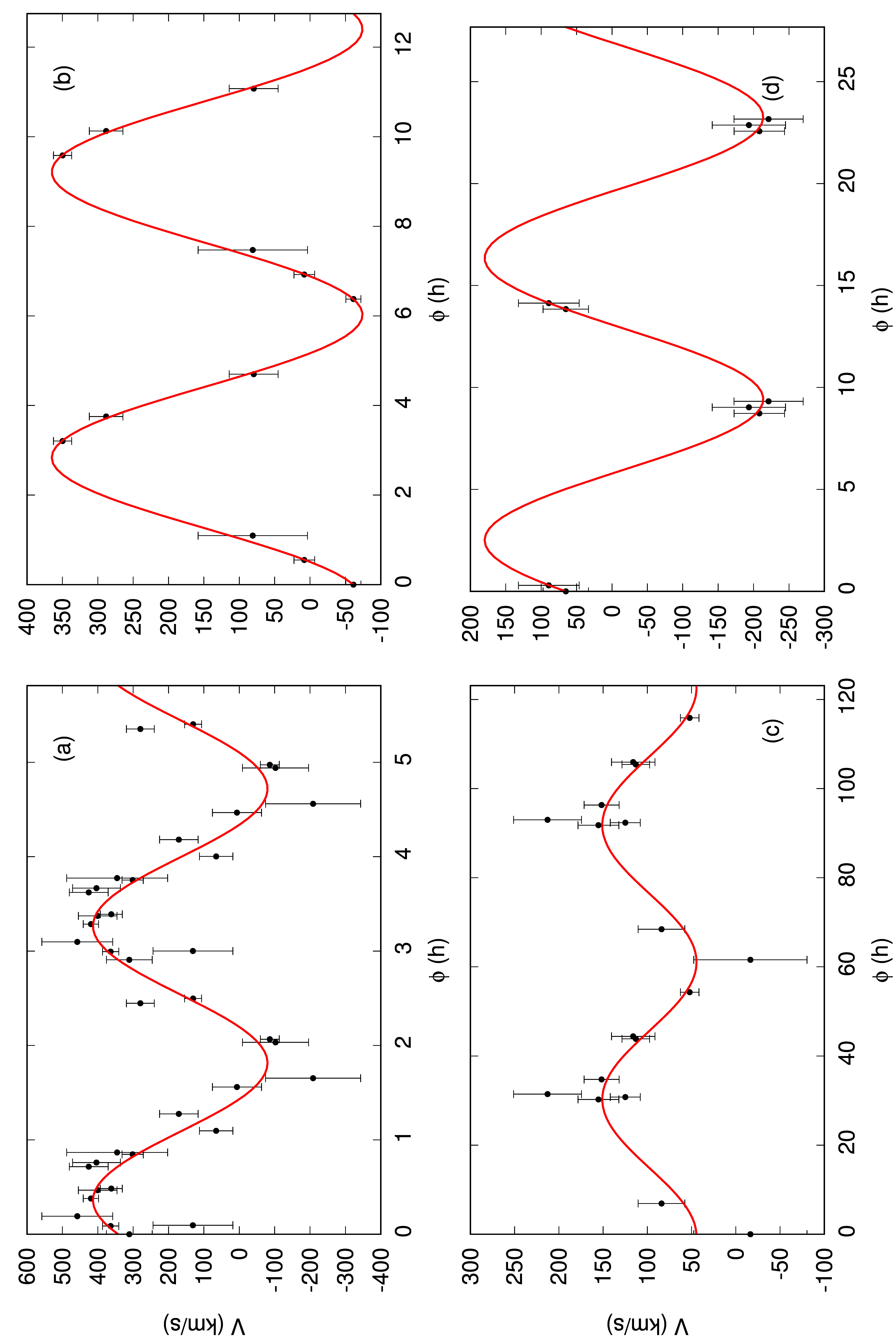}
\caption{Best orbital solutions for the four objects for which we were able to constrain the period.}
\label{SDSS_ELMs_orb}
\end{figure}

\begin{figure}
\includegraphics[angle=-90,width=0.5\textwidth]{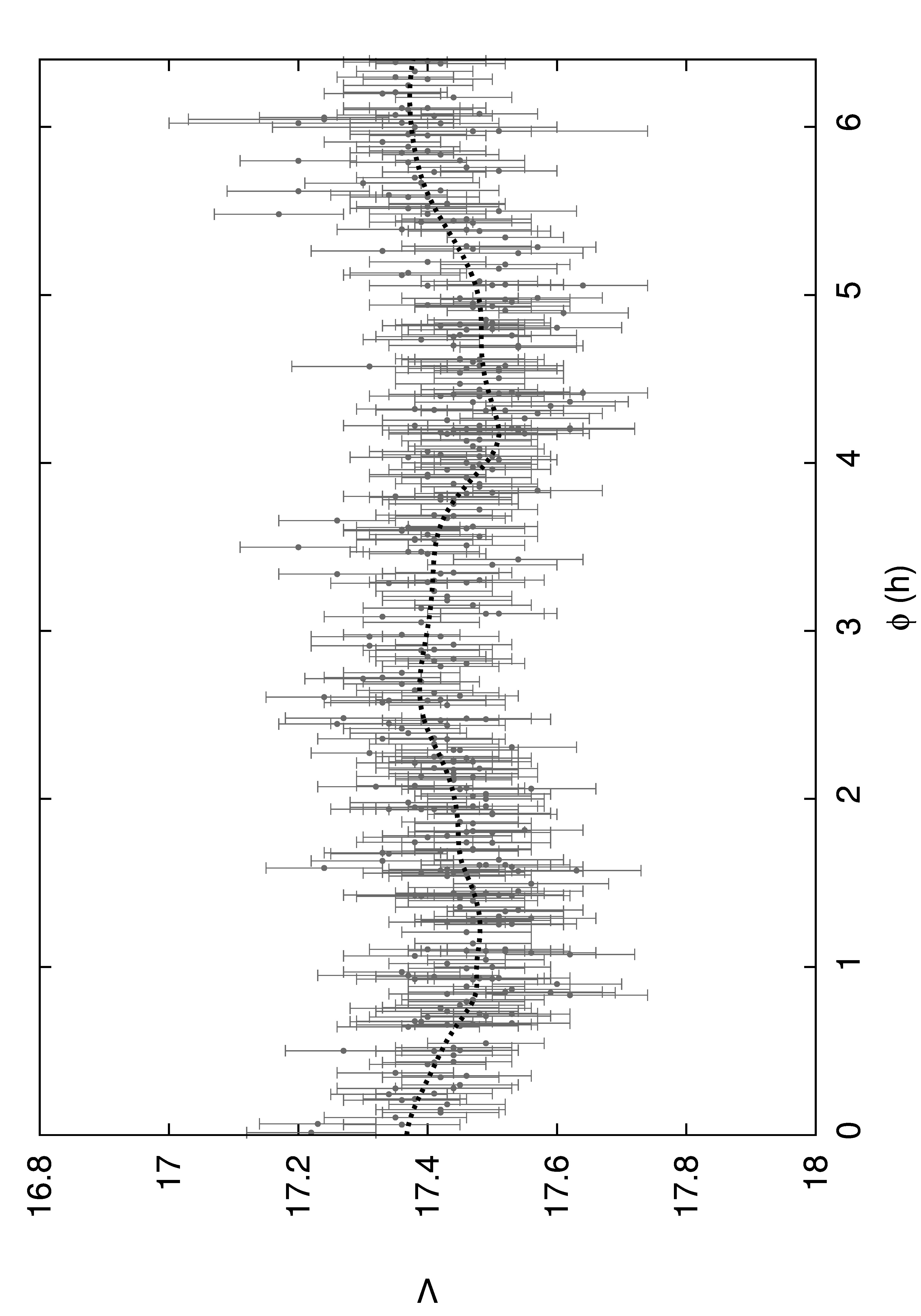}
\caption{CSS light curve for SDSS~J120616.93+115936.2, phase-folded to the 6.4~h, which is the same obtained analysing the velocities, suggesting the variability is due to either eclipses or ellipsoidal variation.}
\label{lc}
\end{figure}

Using the radial velocities estimated from the SDSS spectra of these objects, we attempted to obtain their orbital parameters. The best obtained results are shown on Table \ref{tab_SDSS_ELMs}. The best orbital solutions are shown on Fig. \ref{SDSS_ELMs_orb}. SDSS~J104826.86-000056.7 has nineteen subspectra, which were enough to constrain the period and obtain a good orbital solution. SDSS~J120616.93+115936.2 has only seven subspectra, but its light curve on the Catalina Sky Survey (CSS) shows variability with a period which was consistent with the highest peak on the Fourier transform of the velocities. The phase-folded light curve is shown in Fig. \ref{lc}. SDSS~J045947.40-040035.2 has ten subspectra, but the spacing is such that many aliases arise in the Fourier transform, and in fact periods ranging from 10~h to 60~h had orbital solutions with similar residuals. As previously stated, follow-up is definitely needed to study the nature of this object. SDSS~J171906.23+254142.3 has five subspectra, but less aliasing than SDSS~J045947.40-040035.2, suggesting a period between 8~h and 14~h. We were not able to find a good solution for SDSS~J122911.49-003814.4, which has six subspectra, therefore follow-up is required to probe its nature.

Out of these five objects, we conclude that SDSS~J104826.86-000056.7 and SDSS~J120616.93+115936.2 are unarguably new ELMs, given that both their atmospheric and orbital parameters are consistent with the class. The three remaining objects show the solar abundance $\log g < 5.0$. SDSS~J122911.49-003814.4, however, has $\log g > 5.0$ when the pure-hydrogen models are used. Its spectra does not show strong metal lines, so it is a good ELM candidate. The confirmation of its nature is pending on follow-up studies that can allow the determination of its orbital parameters. SDSS~J171906.23+254142.3 still shows $\log g < 5.0$ on the pure-hydrogen models, but the obtained radial velocity amplitude (197~km/s) can only be explained if the object is in a close binary, requiring it to be compact, therefore it is most likely an ELM. The most uncertain object is SDSS~J045947.40-040035.2, which has $\log g$ in the threshold between main sequence and ELM assuming both models. The estimated distance assuming a main sequence radius is 16~kpc, and its velocities are consistent with the halo. The obtained period and amplitude are also consistent with a main sequence object. Given all that, SDSS~J045947.40-040035.2 is probably a blue straggler star in the halo.

\section{Discussion}

We analysed O, B and A type stars identified by the SDSS pipeline, and estimated their $T\eff$ and $\log g$ from their spectra, using spectral models derived from solar abundance atmospheric models. Comparing the results to our pure-hydrogen solutions published in Kepler et al. (2016), we showed that the addition of metals causes a shift in $\log g$ that is downwards for objects with $\log g > 4.5$, but upwards for objects with $\log g < 4.5$. No general conclusion can be made as to whether the pure-hydrogen models are in fact overestimating the $\log g$, as was suggested by Brown et al. (2017), since the correction depends on the $\log g$. Moreover, although some objects do show many metallic lines in their spectra, others are restricted to Ca and Mg, which are also seen in the known ELMs, due to the fact that rotation has the power to counteract the gravitational settling (Istrate et al. 2016). It is clear from these studies that neither of these two grid of models are in fact adequate, they provide only rough estimates on the parameters, which are dependent on the metallicity. These spectra need to be analysed with more general grids, spanning different metallicities.

Independent of the estimated $\log g$, the magnitudes of the objects suggest that they can not simply be main sequence objects. Assuming a main sequence radius, we estimate distances which are not consistent with a disk distribution. The velocities are also not consistent with the disk and not even with the halo, with over 30 per cent of the O, B and A objects showing velocities more than 4-$\sigma$ above the halo mean velocity. The most probable reason is that the radius estimate --- assuming that the objects have main sequence radius ---  is wrong. If we assume they are He core objects, pre-ELMs and ELMs, they show a distribution consistent with the disk. Another possibility is that the high proper motion and estimated high radial velocity, leading to high spatial velocities, are actually due to orbital motion. They could be blue stragglers in the halo. Models by Schneider et al. (2015) suggest that mass accretion can make a star appear up to 10 times younger than its parent population, which would be sufficient to make an A star survive long enough in the halo. This is in agreement with previous studies in the literature, which find that $\sim 50$~per cent of stars with A-type spectra in the halo are presumably blue stragglers (e.g. Norris \& Hawkins 1991, Kinman, Suntzeff, \& Kraft 1994, Preston, Schectman \& Beers 1994, Clewley et al. 2004, Brown et al. 2008, Xue et al. 2008). The remaining objects are mostly explained as BHBs, hence He-core stars. Some authors suggest that a few could in fact be main sequence stars with an extragalactic origin to explain their young ages (e.g. Preston, Schectman \& Beers 1994). The (pre-)ELM explanation is mostly ignored by these studies, since this is a relatively new class. The sdAs could also be binaries of a hot subdwarf with a main sequence star, but the UV colours suggest that this is not the case, since they do not show significant flux in the UV.%%rejected here

Our most significant result is that the sdAs are clearly composed of two populations. One population contains the red objects, and it has no overlap with the known ELMs. On the other extreme, there is a blue population, which does overlap with known ELMs, but contains cooler objects. The red distribution is possibly dominated by metal-poor main sequence late-type stars, which can be found in the halo, with contamination of cooler pre-ELMs and ELMs, since there is an intersection with the blue distribution. The blue distribution, on the other hand, should contain the missing cool pre-ELM and ELM population, which is under-represented in the literature. Evolutionary models predict that ELMs spend about the same amount of time above and below $T\eff = 8\,500$~K; however, their cooling time-scale is dictated by residual burning. On one hand, this time-scale can be prolonged if mass loss is not effective, so that the star is left with a thick hydrogen atmosphere, where burning via $p-p$ chain reaction will occur (e.g. Maxted et al. 2014). On the other hand, instead of a smooth transition from pre-ELM to ELM, the star can undergo episodes of unstable CNO burning, or shell flashes, that shorten the cooling time-scale by reducing the hydrogen mass on the surface (Althaus et al. 2013, Istrate et al. 2016). As there are many uncertainties in the models, concerning e.g. assumptions on element diffusion, progenitor initial mass and metallicity, and rotation, the cooling time scale between models can vary by more than a factor of two. Brown et al. (2017) estimated a 1:2 ratio of ELMs in the ranges $6500 < T_{\textrm{\tiny eff}} < 9000$~K to $10\,000 < T_{\textrm{\tiny eff}} < 15\,000$~K. Propagating the factor of two uncertainty in the cooling time scale, these ratio can be from 1:4 to 1:1, so 20--50 per cent of the ELMs should show $T\eff < 9\,000$~K; however, as a systematic effect of the search criterion, less than 5 per cent of the published ELMs are in this range. Moreover, the ratio of $\log g = 6-7$ to $\log g = 5-6$ is about 3:4 in the ELM survey, totally dominated by selection effects, while the brightness difference suggests it should be 1:100.

Analysing the SDSS radial velocities, we confirm two new ELMs, SDSS~J104826.86-000056.7 and SDSS~J120616.93+115936.2. SDSS~J120616.93+115936.2 also shows photometric variability with the same period as the orbital period. Two other objects are most likely ELMs. SDSS~J171906.23+254142.3, although showing $\log g < 5.0$, has an amplitude of almost 200~km/s in its best orbital fit. However, as only five subspectra are available, the period is not well constrained, and follow-up should be done to confirm the nature of this object. The SDSS subspectra of SDSS~J122911.49-003814.4 did not allow the estimate of its period, but the high amplitude between its subspectra and its $\log g$ above the main sequence limit favour the ELM classification. All of these objects show $T\eff < 9000$~K. There are only six confirmed ELMs in close binaries in this range (Brown et al. 2016), reflecting the lack of effort to find ELMs in the cool end of the distribution, hence the importance of further studying the objects found here. Finally, we also find SDSS~J045947.40-040035.2 most likely to be a blue straggler star in the halo.

Our effort shows that more than one evolutionary channel is definitely needed to explain the sdA population. For one, there are definitely He core objects such as pre-ELMs and ELMs in the sample. Even if only a small percentage of sdAs is confirmed as ELMs, the number would be high enough to significantly increase the number of known ELMs, especially at the cool end of the distribution. Our understanding of binary evolution, and especially of the common envelope phase that ELMs must experience, can be much improved if we have a sample covering all parameters predicted by these models. The sdA sample can provide that. Our understanding of the formation and evolution of the Galactic halo would also benefit from more detailed study of the sdAs. Many seem to be in the halo with ages and velocities not consistent with the halo population. It is possible that accreted stars from neighbouring dwarf galaxies might be among them. Those whose velocities are in fact consistent with the halo can in turn help us study its dynamics and possibly better constrain the gravitational potential of the halo. The key message of our results is that we should not overlook the complexity of the sdAs. They are of course not all pre-ELM or ELM stars, but they cannot be explained simply as main sequence metal-poor A--F stars. They are most likely products of binary evolution and as such are a valuable asset for improving our models.%add something on the papers Warren's mentioned here

%even though the time spent with log g 6-7 is 25 times larger, the volume is 10 times larger for log 5-6, so we should expect a fraction of 1:5.

\section*{References}

\begin{itemize}
\item Althaus, L. G., Miller Bertolami, M. M., Córsico, A. H. {\bf 2013}, \textit{A\&A}, 557, id. A19, 12 pp.
\item Badenes, C., Maoz, D. {\bf 2012}, \textit{ApJ}, 749, id. L11, 5 pp.
\item Bland-Hawthorn, J., Gerhard, O. {\bf 2016}, \textit{Annual Review of Astronomy and Astrophysics}, 54, 529-596.
\item Brown, W. R., Beers, T. C., Wilhelm, R., Allende-Prieto, C., Geller, M. J., Kenyon, S. J., Kurtz M. J. {\bf 2008}, \textit{ApJ}, 135, 564-574.
\item Brown, W. R., Gianninas, A., Kilic, M., Kenyon, S. J., Allende Prieto, C. {\bf 2016}, \textit{ApJ}, 818, id. 155, 13 pp.
\item Brown, W. R., Kilic, M., Allende Prieto, C., Gianninas, A., Kenyon, S. J. {\bf 2013}, \textit{ApJ}, 769, id. 66, 11 pp.
\item Brown, W. R., Kilic, M., Allende Prieto, C., Kenyon, S. J. {\bf 2010}, \textit{ApJ}, 723, 1072-1081.
\item Brown, W. R., Kilic, M., Allende Prieto, C., Kenyon, Scott J. {\bf 2012}, \textit{ApJ}, 744, id. 142, 12 pp.
\item Brown, W. R., Kilic, M., Gianninas, A. {\bf 2017}, \textit{ApJ}, 839, id. 23, 12 pp.
\item Clewley, L., Warren, S. J., Hewett, P. C., Norris, John E., Evans, N. W. {\bf 2004}, {\it MNRAS}, 352, 285-298.
\item De Rosa, R. J., Patience, J., Wilson, P. A., Schneider, A., Wiktorowicz, S. J., Vigan, A. {\bf 2014}, \textit{MNRAS}, 437, 1216-1240.
\item Duchêne G., Kraus A. {\bf 2013}, \textit{ARA\&A}, 51, 269-310.
\item Gentile-Fusillo, N. P., Gaensicke, B. T., Greiss, S. {\bf 2015}, \textit{MNRAS}, 448, 2260-2274.
\item Gianninas, A., Kilic, M., Brown, W. R., Canton, P., Kenyon, S. J. {\bf 2015}, \textit{ApJ}, 812, id. 167, 12 pp.
\item Hermes, J. J., Gänsicke, B. T., Breedt, E. {\bf 2017}, In: P.-E. Tremblay, B. Gaensicke, T. Marsh (Ed.), Proceedings of 20th European White Dwarf Workshop (25-29 July 2016, Coventry, UK), \textit{ASP Conference Series}, 509, 453-459. 
\item Istrate A. G., Marchant P., Tauris T. M., Langer N., Stancliffe R. J., Grassitelli L. {\bf 2016}, \textit{A\&A}, 595, id. A35, 24 pp.
\item Kepler, S. O., Pelisoli, I., Koester, D., Ourique, G., Romero, A. D., Reindl, N., et al. {\bf 2016}, \textit{MNRAS}, 455, 3413-3423.
\item Kilic, M., Brown, W. R., Allende Prieto, C., Agüeros, M. A., Heinke, C., Kenyon, S. J. {\bf 2011}, \textit{ApJ}, 727, id. 3, 12 pp.
\item Kilic, M., Brown, W. R., Allende Prieto, C., Kenyon, S. J., Heinke, C. O., Agüeros, M. A., et al. {\bf 2012}, \textit{ApJ}, 751, id. 141, 13 pp.
\item Kilic, M., Stanek, K. Z., Pinsonneault, M. H. {\bf 2007}, \textit{ApJ}, 671, 761-766.
\item Kinman T. D., Suntzeff N. B., Kraft R. P. {\bf 1994}, {\it AJ}, 108, 1722-1772.
\item Koester, D. {\bf 2010}, \textit{MmSAI}, 81, 921-931.
\item Kordopatis, G., Recio-Blanco, A., de Laverny, P., Gilmore, G., Hill, V., Wyse, R. F. G. {\bf 2011}, \textit{A\&A}, 535, id. A107, 18 pp.
\item Lance, C. M. {\bf 1988}, {\it AJ}, 334, 927-946.
\item Lenz, P., Breger, M. {\bf 2005}, \textit{Communications in Asteroseismology}, 46, 53-136.
\item Maxted P. F. L., Bloemen S., Heber U., Geier S., Wheatley P. J., Marsh T. R., Breedt E., Sebastian D., Faillace G., Owen C., Pulley D., Smith D., Kolb U., Haswell C. A., Southworth J., Anderson D. R., Smalley B., Collier C. A., Hebb L., Simpson E. K., West R. G., Bochinski J., Busuttil R., Hadigal S. 2014, MNRAS, 437, 1681-1697.
\item Munn, J. A., Monet, D. G., Levine, S. E., Canzian, B., Pier, J. R., Harris, H. C. et al. {\bf 2004}, \textit{AJ}, 127, 3034-3042.
\item Munn, J. A., Harris, H. C., von Hippel, T., Kilic, M., Liebert, J. W., Williams, K. A. et al. {\bf 2014}, \textit{AJ}, 148, id. 132, 13 pp.
\item Norris \& Hawkins 1991
\item Pelisoli, I., Kepler, S. O., Koester, D., Romero, A. D. {\bf 2017}, In: P.-E. Tremblay, B. Gaensicke, T. Marsh (Ed.), Proceedings of 20th European White Dwarf Workshop (25-29 July 2016, Coventry, UK), \textit{ASP Conference Series}, 509, 447-452. 
\item Preston, G. W., Beers, T. C., Shectman, S. A. {\bf 1994}, {\it AJ}, 108, 538-554.
\item Rodgers A. W., Harding P., Sadler E. {\bf 1981}, {\it AJ}, 244, 912-918.
\item Romero, A. D., Campos, F., Kepler, S. O. {\bf 2015}, \textit{MNRAS}, 450, 3708-3723.
\item Schneider, F. R. N., Izzard, R. G., Langer, N., de Mink, S. E. {\bf 2015}, \textit{ApJ}, 805, id. 20, 20 pp.
\item Tokovinin A. {\bf 2014}, \textit{AJ}, 147, id. 87, 14 pp. 
\item Xue et al. 2008
\item Yuan, H. B., Liu, X. W., Xiang, M. S. {\bf 2013}, \textit{MNRAS}, 430, 2188-2199.
\end{itemize}	
   
\end{document}